\documentclass[preprint2]{aastex62}

\newcommand\aastex{AAS\TeX}

\usepackage{epstopdf}

\received{March 1, 2018}
\revised{April 1, 2018}
\accepted{\today}

\submitjournal{ApJ}

\shorttitle{\aastex\ Vertical oscillations of a coronal cavity}
\shortauthors{Zhang \& Ji}

\begin{document}

\title{Vertical oscillation of a coronal cavity triggered by an EUV wave}

\correspondingauthor{Q. M. Zhang}
\email{zhangqm@pmo.ac.cn}

\author[0000-0003-4078-2265]{Q. M. Zhang}
\affil{Key Laboratory for Dark Matter and Space Science, Purple Mountain Observatory, CAS, Nanjing 210034, China}
\affil{CAS Key Laboratory of Solar Activity, National Astronomical Observatories, Beijing 100012, China}

\author{H. S. Ji}
\affil{Key Laboratory for Dark Matter and Space Science, Purple Mountain Observatory, CAS, Nanjing 210034, China}

\begin{abstract}
In this paper, we report our multiwavelength observations of the vertical oscillation of a coronal cavity on 2011 March 16.
The elliptical cavity with an underlying horn-like quiescent prominence was observed by the Atmospheric Imaging Assembly (AIA) on board the \textit{Solar Dynamics Observatory} (\textit{SDO}).
The width and height of the cavity are 150$\arcsec$ and 240$\arcsec$, and the centroid of cavity is 128$\arcsec$ above the solar surface.
At $\sim$17:50 UT, a C3.8 two-ribbon flare took place in active region 11169 close to the solar western limb. Meanwhile, 
a partial halo coronal mass ejection (CME) erupted and propagated at a linear speed of $\sim$682 km s$^{-1}$.
Associated with the eruption, a coronal extreme-ultraviolet (EUV) wave was generated and propagated in the northeast direction at a speed of $\sim$120 km s$^{-1}$. 
Once the EUV wave arrived at the cavity from the top, it pushed the large-scale overlying magnetic field lines downward before bouncing back. 
At the same time, the cavity started to oscillate coherently in the vertical direction and lasted for $\sim$2 cycles before disappearing. 
The amplitude, period, and damping time are 2.4$-$3.5 Mm, 29$-$37 minutes, and 26$-$78 minutes, respectively. 
The vertical oscillation of the cavity is explained by a global standing MHD wave of fast kink mode. 
To estimate the magnetic field strength of the cavity, we use two independent methods of prominence seismology. 
It is found that the magnetic field strength is only a few Gauss and less than 10 G.
\end{abstract}


\section{Introduction} \label{sec:intro}
Solar prominences are cool and dense plasma structures suspending in the corona \citep{lab10,mac10,par14}. The routine observation of prominences has a long history. 
They can be observed in radio, H$\alpha$, Ca\,{\sc ii}\,H, He\,{\sc i}\,10830 {\AA}, extreme-ultraviolet (EUV), and soft X-ray (SXR) wavelengths \citep[e.g.,][]{gop03,ji03,ber08,zqm15,ste15,wang16}.
When prominences appear on the disk in the H$\alpha$ full-disk images, they look darker than the surrounding area due to the absorption of the background radiation.
Therefore, they are also called filaments \citep{mar98}. Prominences (or filaments) can form in active regions (ARs), quiet region, and polar region \citep[e.g.,][]{hei08,su15,yan15}.
It is generally believed that the gravity of prominences should be balanced by the upward magnetic tension force of the dips in sheared arcades \citep{jor92,mar01,xia12} or 
magnetic flux ropes \citep[MFRs;][]{aul98,hil13,xia16,yan18}.

Prominences tend to oscillate after being disturbed \citep[][and references therein]{oli92,tri09,arr12,luna18}. 
According to the amplitude, prominence oscillations can be classified into the small-amplitude \citep[e.g.,][]{oka07,ning09} and large-amplitude \citep{chen08} types. 
According to the period, prominence oscillations can be divided into the short-period ($\leq$10 min) and long-period ($\geq$40 min) \citep{lin07,zqm17a} types. 
A more reasonable classification is based on the direction of oscillation. For longitudinal oscillations, the prominence
material oscillates along the threads with small angles of $10^{\circ}-20^{\circ}$ between the threads and spine \citep{jing03,zqm12,li12,luna12,luna14}.
For transverse oscillations, the direction is horizontal (vertical) if the whole body of the prominence oscillates parallel (vertical) to the solar surface \citep{hyd66,ram66,kle69,shen14a}.
Generally speaking, the amplitudes and periods of longitudinal oscillations are larger than those of transverse oscillations. 
Occasionally, two types of oscillations coexist in a single event, showing a very complex behavior \citep{gil08,shen14b,zqm17b}. 
Recently, state-of-the-art magnetohydrodynamics (MHD) numerical simulations have thrown light on the triggering mechanisms, restoring forces, and damping mechanisms of prominence
oscillations \citep{zqm13,ter13,ter15,luna16,zhou17,zhou18}.

Coronal cavities are dark structures as a result of density depletion in the white light (WL), EUV, and SXR wavelengths \citep{vai73,gib06,vas09}. 
The densities of cavities are 25\%$-$50\% lower than the adjacent streamer material \citep{mar04,ful08,ful09,sch11}. 
Cavities show diverse shapes, such as semi-circular, circular, elliptical, and teardrop \citep{for13,kar15a}. 
The lengths, heights, widths of cavities are 0.06$-$2.9 $R_{\odot}$, 0.08$-$0.5 $R_{\odot}$, and 0.09$-$0.4 $R_{\odot}$, respectively \citep{kar15a}. 
With a broader differential emission measure distribution, the average temperatures (1$-$2 MK) of cavities are slightly higher than the streamers \citep{hud99,hab10,ree12}. 
Both observations and numerical experiments indicate that the magnetic structure of cavities could be excellently described by a twisted/helical MFR \citep{gib10,dove11,fan12,bak13,rach13,chen18}. 
At the bottom of cavities, there are denser filament plasmas drained down by gravity \citep{low95,reg11}. Spectroscopic observations reveal that there are continuous whirling/spinning motion \citep{wang10} 
and large-scale flows with Doppler speeds of 5$-$10 km s$^{-1}$ in prominence cavities \citep{sch09}. The stable structures can last for days or even weeks to months \citep{kar15b}. 
After being strongly disturbed or a certain type of MHD instability is activated, a cavity may experience loss of equilibrium and erupt to drive a coronal mass ejection \citep[CME;][]{ill85}.

For the first time, \citet{liu12} reported the transverse (horizontal) oscillation of a twisted MFR cavity and its embedded filament after the arrival of the leading EUV wave front at a speed of $\geq$1000 km s$^{-1}$.
The average amplitude, initial velocity, period, and \textit{e}-folding damping time are $\sim$2.3 Mm, $\sim$8.8 km s$^{-1}$, $\sim$27.6 minutes, and $\sim$119 minutes, respectively.
Sometimes, MFRs carrying prominence material undergo global vertical oscillations with periods of hundreds of seconds, which are explained by the standing wave of fast kink mode \citep{kim14,zhou16}.
However, the vertical oscillation of a cavity as a result of the interaction between a coronal EUV wave and the cavity has rarely been observed and investigated. 

In this paper, we report our multiwavelength observations of the vertical oscillation of a coronal cavity triggered by an EUV wave on 2011 March 16.
The wave was caused by the eruption of a C3.8 two-ribbon flare and a partial halo CME at the remote AR 11169. 
The paper is structured as follows. Data analysis is described in detail in Section~\ref{sec:data}. Results are shown in Section~\ref{sec:result}. 
Estimation of the magnetic field strength of the cavity is arranged in Section~\ref{sec:discuss}. Finally, we give a brief summary in Section~\ref{sec:summary}.

\section{Instruments and data analysis} \label{sec:data}
Located north to NOAA AR 11169 (N17W74), the coronal cavity (N62W75) with an underlying prominence was continuously observed by the Global Oscillation Network Group (GONG) in H$\alpha$ line center (6562.8 {\AA})
and by the Atmospheric Imaging Assembly \citep[AIA;][]{lem12} on board the \textit{Solar Dynamics Observatory} \citep[\textit{SDO};][]{pes12} in UV (1600 {\AA}) and EUV (171, 193, 211, and 304 {\AA}) wavelengths. 
The photospheric line-of-sight (LOS) magnetograms were observed by the Helioseismic and Magnetic Imager \citep[HMI;][]{sch12} on board \textit{SDO}. 
The level\_1 data from AIA and HMI were calibrated using the standard \textit{Solar Software} (\textit{SSW}) programs \textit{aia\_prep.pro} and \textit{hmi\_prep.pro}. 
The full-disk H$\alpha$ and AIA 304 {\AA} images were coaligned with a precision of $\sim$1$\farcs$2 using the cross correlation method. 
The CME was observed by the C2 WL coronagraph of the Large Angle Spectroscopic Coronalgraph \citep[LASCO;][]{bru95} on board \textit{SOHO}\footnote{http://cdaw.gsfc.nasa.gov/CME\_list/}.
The LASCO/C2 data were calibrated using the \textit{SSW} program \textit{c2\_calibrate.pro}. The CME was also observed by the COR1\footnote{http://cor1.gsfc.nasa.gov/catalog/cme/2011/} 
with a field of view (FOV) of 1.3$-$4.0 $R_{\odot}$ on board the \textit{Solar TErrestrial RElations Observatory} \citep[\textit{STEREO};][]{kai05}.
The ahead satellite (hereafter STA) and behind satellite (hereafter STB) had separation angles of $\sim$88$^{\circ}$ and $\sim$95$^{\circ}$ with respect to the Sun-Earth direction on 2011 March 16.
The coronal EUV wave associated with the CME was observed by AIA and the Extreme-Ultraviolet Imager (EUVI) in the Sun Earth Connection Coronal and Heliospheric Investigation package \citep[SECCHI;][]{how08}.
EUVI observes the Sun in four wavelengths (171, 195, 284, and 304 {\AA}).
Calibrations of the COR1 and EUVI data were performed using the \textit{SSW} program \textit{secchi\_prep.pro}. The deviation of STA north-south direction from the solar rotation axis was corrected.
The EUV flux of the C3.8 flare in 1$-$70 {\AA} was recorded by the Extreme Ultraviolet Variability Experiment \citep[EVE;][]{wood12} on board \textit{SDO}.
The SXR flux of the flare in 1$-$8 {\AA} was recorded by the \textit{GOES} spacecraft.
To investigate the hard X-ray (HXR) source of the flare, we made HXR images at different energy bands (6$-$12 keV and 12$-$25 keV) observed by the \textit{Reuven Ramaty High-Energy Solar Spectroscopic Imager} 
\citep[\textit{RHESSI};][]{lin02}. The HXR images were generated using the CLEAN method with integration time of 10 s.
The observational parameters, including the instrument, wavelength, time, cadence, and pixel size are summarized in Table~\ref{tab:para}.

\begin{deluxetable}{ccccc}
\tablecaption{Description of the observational parameters \label{tab:para}}
\tablecolumns{5}
\tablenum{1}
\tablewidth{0pt}
\tablehead{
\colhead{Instru.} &
\colhead{$\lambda$} &
\colhead{Time} & 
\colhead{Caden.} & 
\colhead{Pix. size} \\
\colhead{} & 
\colhead{({\AA})} &
\colhead{(UT)} & 
\colhead{(s)} & 
\colhead{(\arcsec)}
}
\startdata
  LASCO & WL & 19:00$-$23:36 & 720 & 11.4 \\
  COR1   & WL & 19:00$-$20:30 & 300 & 15.0 \\
  HMI &  6173        & 17:30$-$23:30 & 45 & 0.6 \\
  AIA & 1600 & 17:30$-$23:30 & 24 & 0.6 \\
  AIA & 171$-$304 & 17:30$-$23:30 & 12 & 0.6 \\
  EUVI & 171       & 17:30$-$23:30 & 150 & 1.6 \\
  EUVI & 195, 284       & 17:30$-$23:30 & 300 & 1.6 \\
  GONG & 6563 & 17:30$-$23:30 & 60  & 1.0 \\
\textit{GOES}  & 1$-$8  & 17:30$-$23:30 & 2.05  & \nodata \\
    EVE & 1$-$70  & 17:30$-$23:30 & 0.25 & \nodata \\
\textit{RHESSI} & 6$-$25 keV & 17:30$-$23:30 & 10 & 4 \\
\enddata
\end{deluxetable}

\section{Results} \label{sec:result}

\subsection{Flare and CME} \label{s-flare}
\begin{figure*}
\plotone{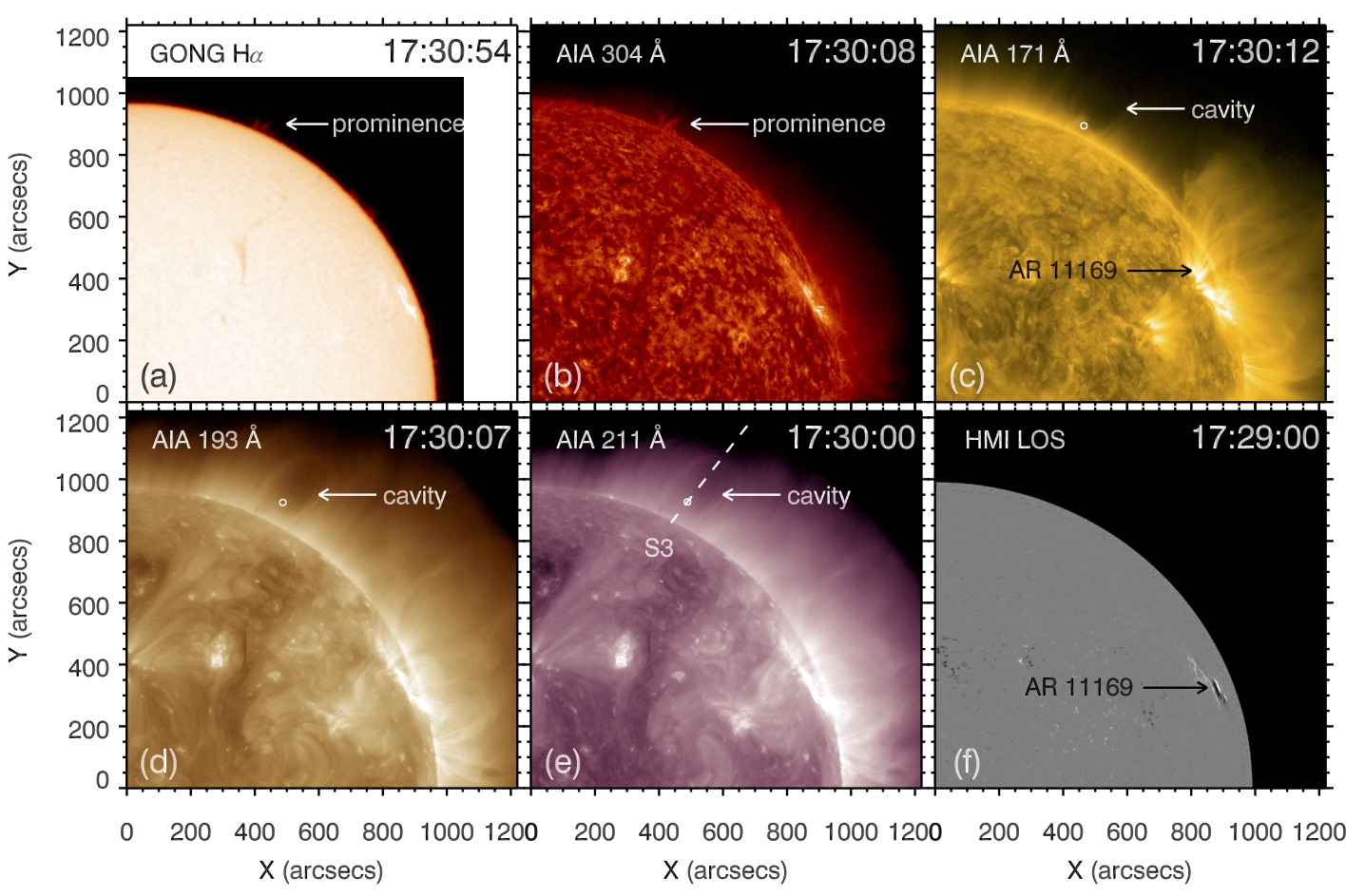}
\caption{(a-e) H$\alpha$ and EUV images at 17:30 UT before the flare. In panels (a-b), the white arrows point at the horn-like quiescent prominence. 
In panels (c-e), the white arrows point at the elliptical dark cavity with depleted EUV emissions. The small white circles mark the initial positions ($y_0$ in Table~\ref{tab:fitting}) of 
the vertical oscillation. The inclined dashed line denotes a selected slice (S3) to investigate the oscillation. (f) HMI LOS magnetogram at 17:29 UT.
\label{fig1}}
\end{figure*}

Figure~\ref{fig1} shows the H$\alpha$ and EUV images before the flare. In panels (a) and (b), the arrows point at the quiescent prominence resembling a horn structure. 
Above the prominence, there is a dark void with depleted EUV emissions (see panels (c-e)). This is the typical coronal cavity showing an elliptical shape. 
The height and width of the cavity are $\sim$240$\arcsec$ and $\sim$150$\arcsec$, and the centroid of the cavity has a height of $\sim$128$\arcsec$, respectively. 
In panel (f), the HMI LOS magnetogram features AR 11169 close to the western solar limb. The distance between the AR and cavity is $\sim$300$\arcsec$.

\begin{figure}
\plotone{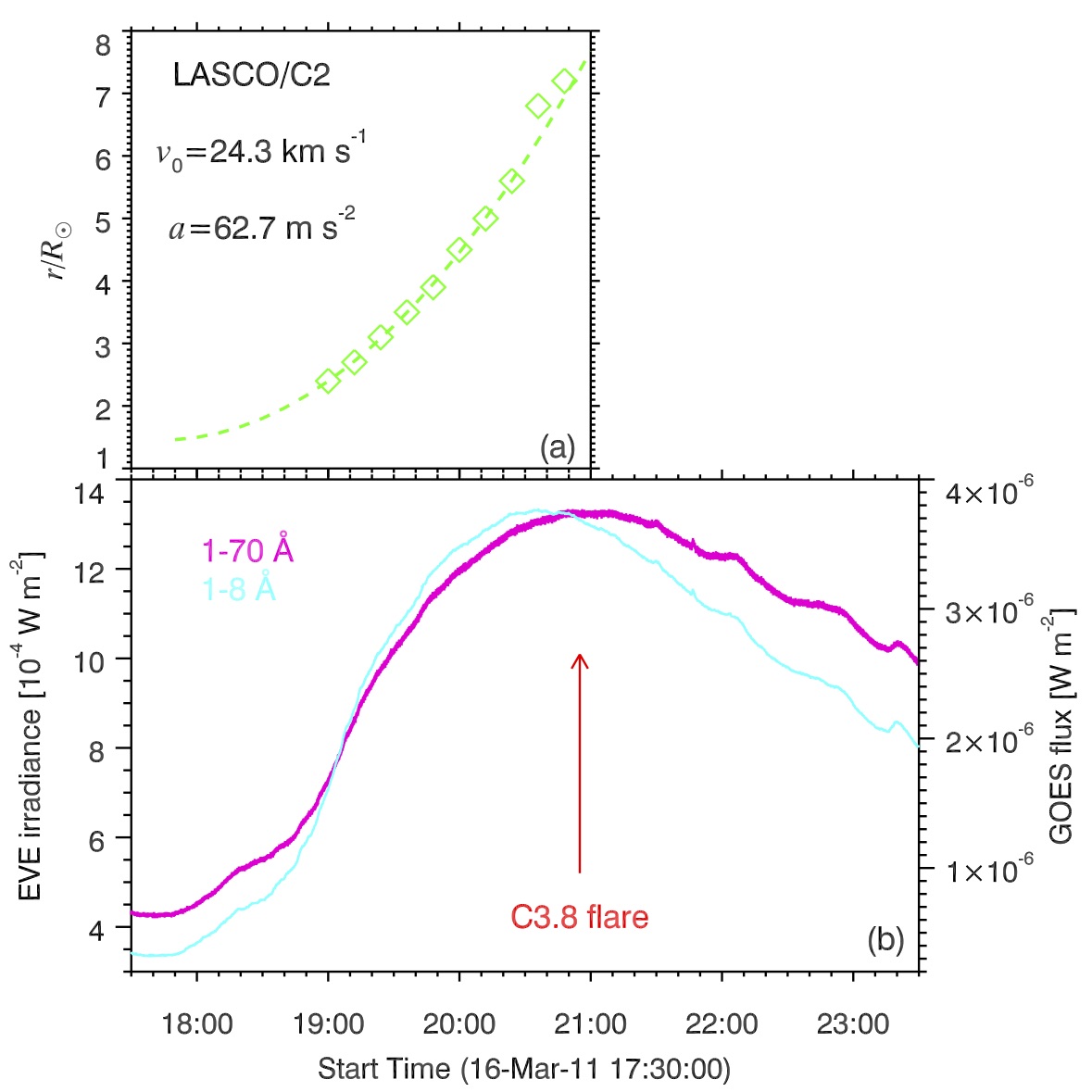}
\caption{(a) Height-time plot of the CME in the FOV of LASCO/C2. A quadratic polynomial is applied to fit the temporal evolution of the height, 
which is represented by the green dashed line. The initial velocity and acceleration are labeled. 
(b) EUV irradiance (magenta line) and SXR flux (cyan line) of the flare recorded by \textit{SDO}/EVE and \textit{GOES}.
\label{fig2}}
\end{figure}

In Figure~\ref{fig2}, the bottom panel shows the temporal evolutions of the EUV irradiance and SXR flux of the flare. It is clear that the EUV and SXR intensities of the flare started to rise 
gradually at $\sim$17:50 UT and reached the peak values around 21:00 UT, which was followed by a long main phase. 

\begin{figure*}
\plotone{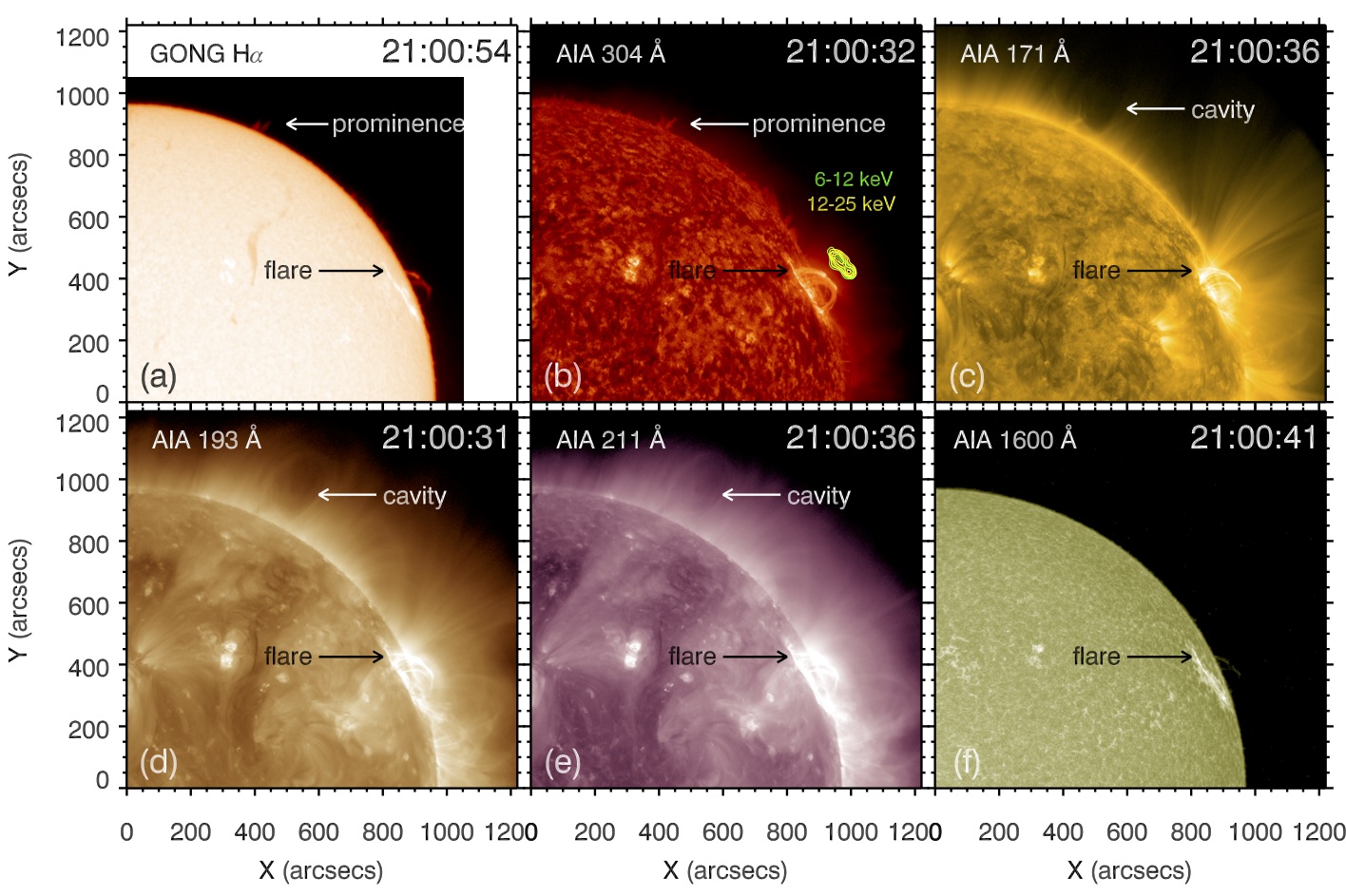}
\caption{H$\alpha$, EUV, and UV images of the flare at 21:00 UT near the flare peak time. The white arrows point at the prominence and cavity. 
The black arrows point at the hot and bright post flare loops. In panel (b), the green and yellow lines represent the intensity contours of the HXR 
emissions at 6$-$12 and 12$-$25 keV energy bands, respectively. (An animation of a combination of panels (c-e) of this figure is available.)
\label{fig3}}
\end{figure*}

The flare observed in H$\alpha$, EUV, and UV wavelengths near the peak time are displayed in Figure~\ref{fig3}. An online animation (\textit{20110316.mov}) is 
a combination of panels (c-e) observed by \textit{SDO}/AIA. With a time cadence of 240 s, the animation starts from 17:44 UT on 2011 March 16 to 00:00 UT on 2011 March 17, 
featuring the vertical oscillation of the dark cavity with an embedded prominence.
The flare is a typical two-ribbon flare with bright ribbons and semi-circular post flare loops in AR 11169. 
In panel (b), an above-the-loop-top HXR source is successfully reconstructed at 6$-$12 keV and 12$-$25 keV energy bands.
The intensity contours of the source are drawn with green and yellow lines.

\begin{figure*}
\plotone{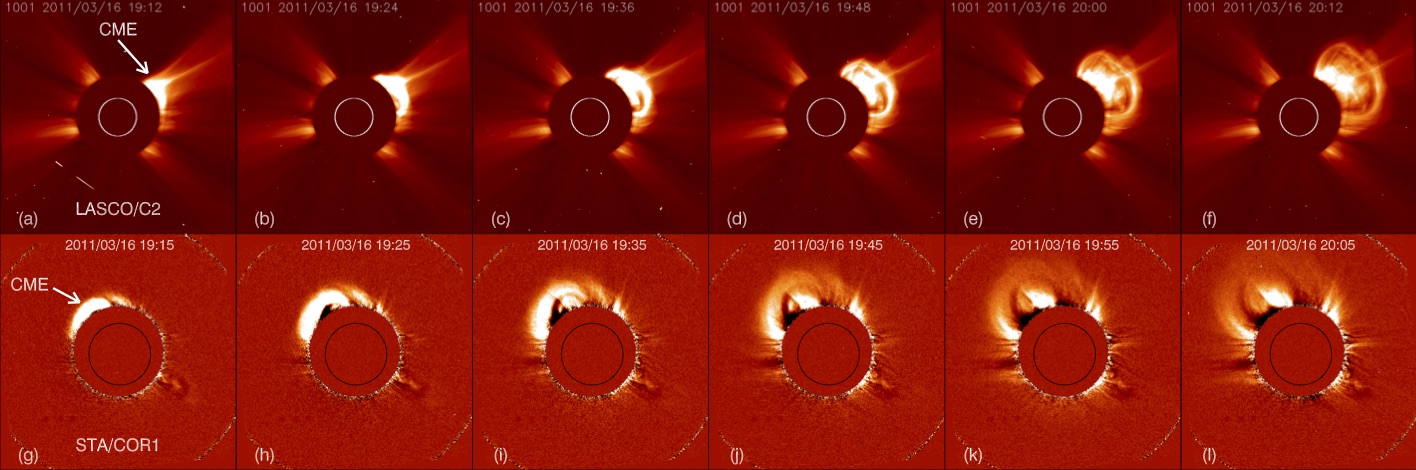}
\caption{WL images of the flare-related partial halo CME in the FOV of LASCO/C2 (top panels) and STA/COR1 (bottom panels).
The arrows in panels (a) and (g) point at the leading edges of the CME.
\label{fig4}}
\end{figure*}

In Figure~\ref{fig4}, the top and bottom panels demonstrate the WL images of the flare-related CME in the FOVs of LASCO/C2 and STA/COR1, respectively. With the central position angle (CPA) and angular width 
being 268$^{\circ}$ and 184$^{\circ}$, the CME appeared first in the FOV of C2 at $\sim$19:12 UT. During its forward propagation at a linear speed of $\sim$682 km s$^{-1}$, 
the CME underwent lateral expansion and evolved into a typical three-part structure, which consists of a bright leading edge, a dark cavity, and a bright core \citep{ill85}.
Evolution of the CME in the FOV of STA/COR1 is similar to that in C2, except for a different CPA due to the different perspective of STA. 

In Figure~\ref{fig2}(a), we plot the temporal evolution of the CME height with green diamonds. A quadratic polynomial ($h=h_{0}+v_{0}t+at^2/2$) is applied to fit the temporal evolution of the CME height, 
which is represented by the green dashed line. The initial velocity $v_{0}=24.3$ km s$^{-1}$ at 17:50 UT and the constant acceleration $a=62.7$ m s$^{-2}$.

\subsection{Coronal EUV Wave} \label{s-wave}

\begin{figure*}
\plotone{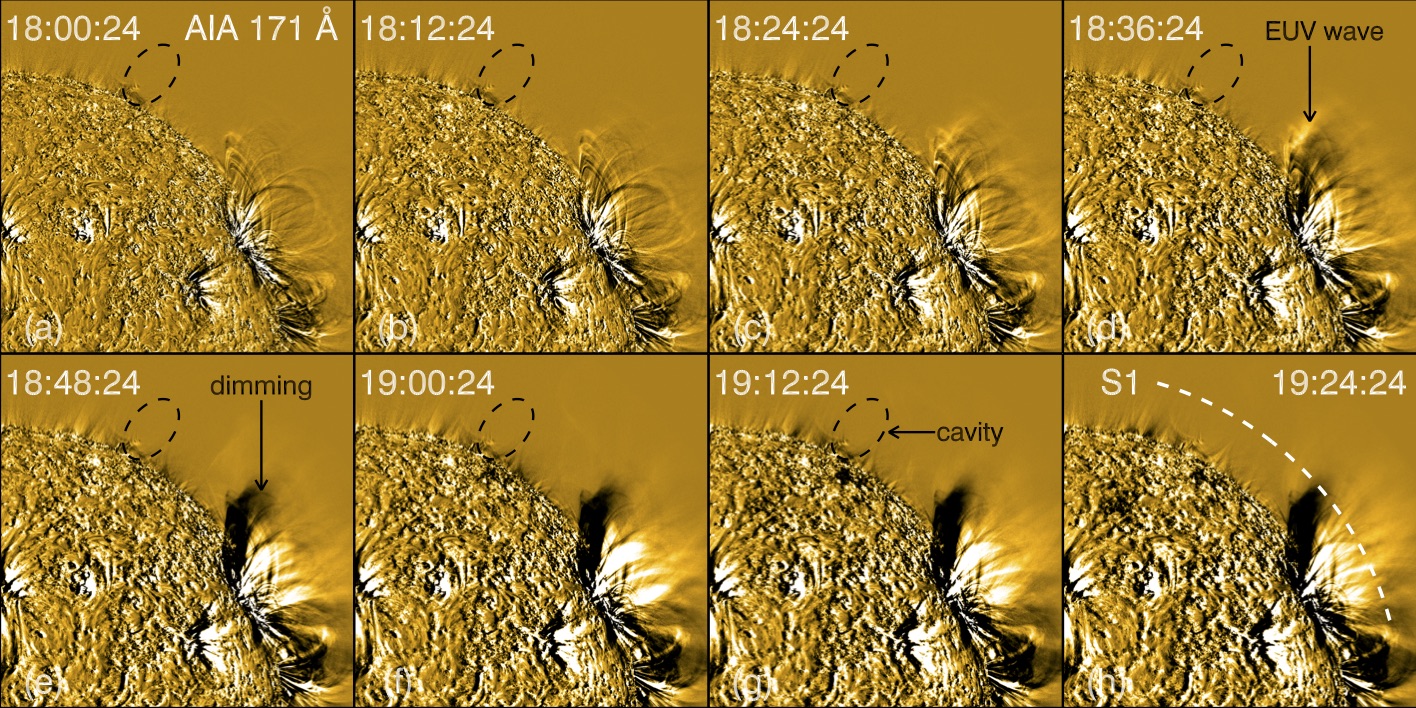}
\caption{Eight snapshots of the base-difference images in AIA 171 {\AA} during 18:00$-$19:24 UT. The black dashed ellipses in each panel denote the position of coronal cavity.
In panel (d), the arrow points at the EUV wave front. In panel (e), the arrow points at the dimming region behind the EUV wave.
In panel (h), the white dashed line denotes a slice (S1) to investigate the evolution of the EUV wave.
\label{fig5}}
\end{figure*}

The fast and wide CME generated a coronal EUV wave propagating on the solar surface. To illustrate the wave more clearly, we take the EUV images at 17:30 UT as base images and obtain base-difference 
images at the following times. Figure~\ref{fig5} demonstrate eight snapshots of the base-difference images in AIA 171 {\AA} during 18:00$-$19:24 UT. It is seen that as the CME proceeds, a bright EUV wave front forms 
when the lateral magnetic field lines of the CME are stretched and pressed, which is indicated by the arrow in panel (d). Behind the EUV wave front, there is a dark dimming region where the electron 
density decreases significantly after the impulsive expulsion of material carried by the CME \citep{zqm17c}.

\begin{figure}
\plotone{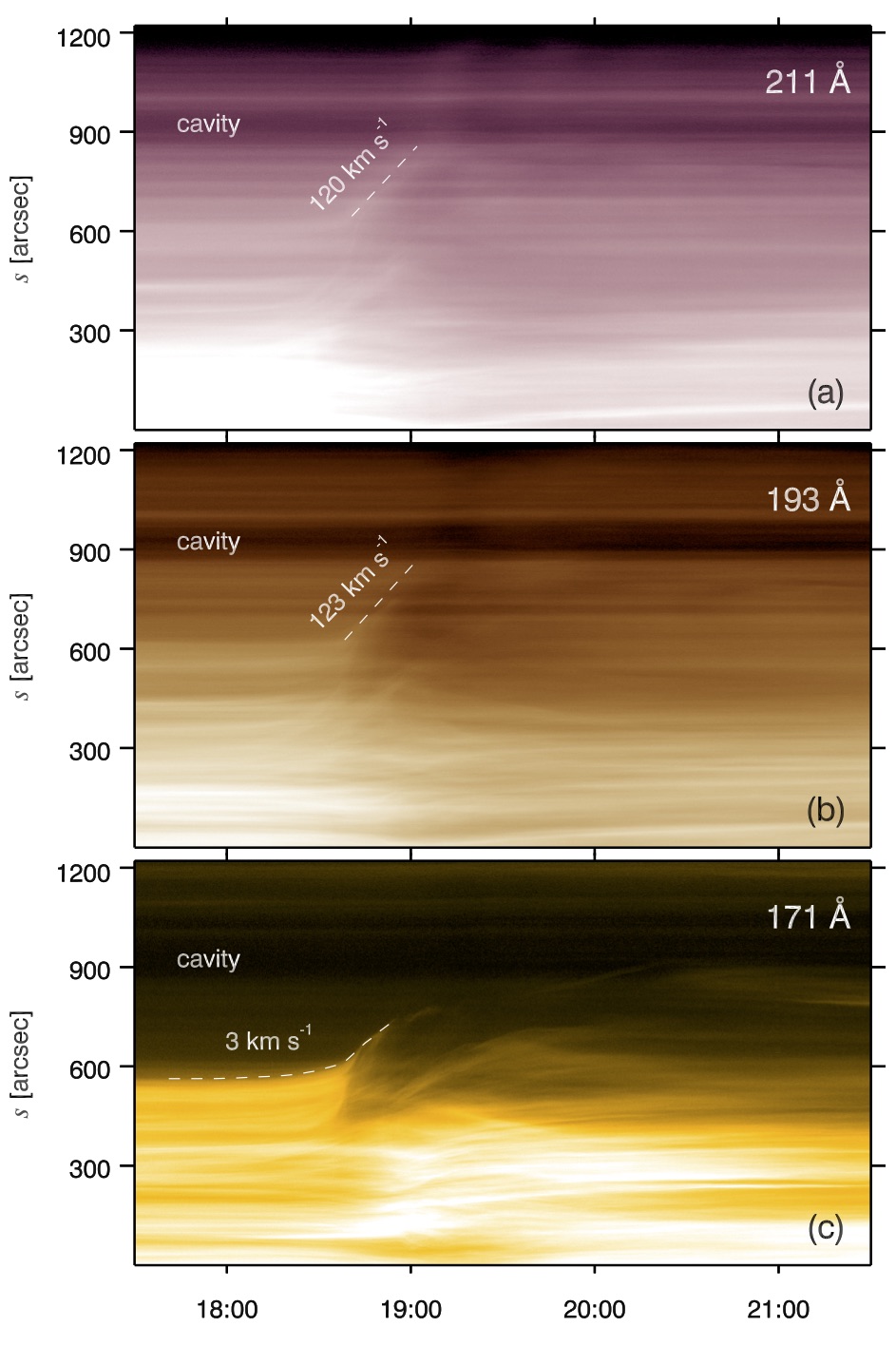}
\caption{Time-slice diagrams of S1 in 211, 193, and 171 {\AA}. $s=0$ and $s=1220\arcsec$ in the $y$-axis stand for the southwest and northeast endpoints of S1.
The velocities ($\sim$120 km s$^{-1}$) of the EUV wave are labeled.  
\label{fig6}}
\end{figure}

In order to investigate the evolution of the EUV wave, we select a curved slice (S1) in Figure~\ref{fig5}(h). The long slice with a length of 1220$\arcsec$ starts from AR 11169 and passes through the cavity. 
Concentric with the solar limb, S1 has a height of 201$\arcsec$. Figure~\ref{fig6} shows the time-slice diagrams of S1 in 211, 193, and 171 {\AA}. 
In panels (a) and (b), the bright inclined feature indicates the propagation of EUV wave in the northeast direction during 18:40$-$19:00 UT, which is overlaid by white dashed lines. 
The slopes of the dashed lines are equal to the velocities ($\sim$120 km s$^{-1}$) of EUV wave. It is obvious that the EUV wave reached and interacted with the dark cavity.
In panel (c), the dashed line indicates the slow lateral expansion of the CME at a speed of $\sim$3 km s$^{-1}$ before 18:40 UT in 171 {\AA}.

\begin{figure*}
\plotone{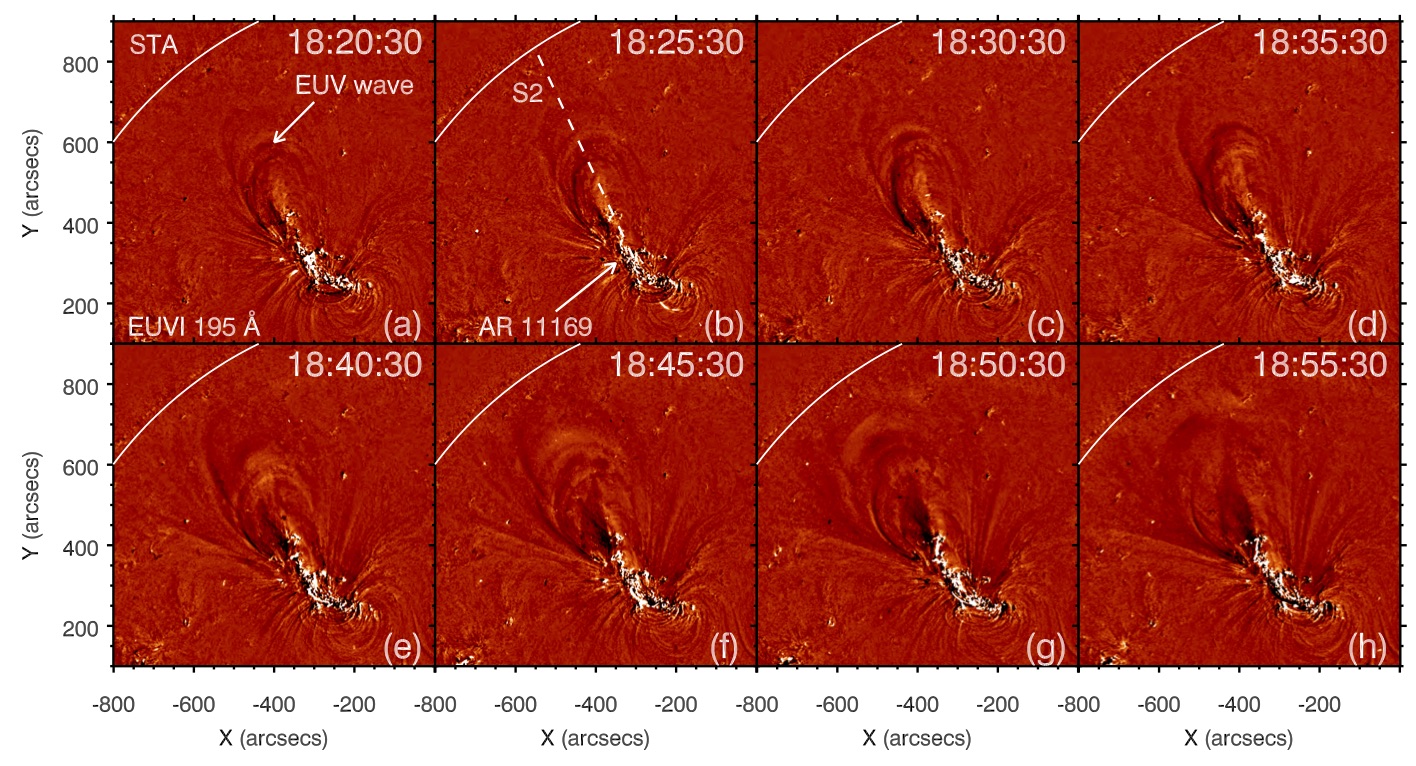}
\caption{Eight snapshots of the base-difference images in EUVI 195 {\AA} during 18:20$-$18:55 UT. The solid curve marks the solar limb in each panel.
In panel (a), the arrow points at the EUV wave front. In panel (b), the arrow points at the AR, and the dashed line (S2) is used to investigate the EUV wave.
\label{fig7}}
\end{figure*}

The EUV wave was also observed by STA/EUVI in various wavelengths. Like in Figure~\ref{fig5}, we take the EUV images at 17:30 UT as base images and obtain base-difference images at the following times.
Eight snapshots of the base-difference images in EUVI 195 {\AA} during 18:20$-$18:55 UT are displayed in Figure~\ref{fig7}. It is clear that as the CME proceeds, the bright EUV wave front propagates in the 
northeast direction. 

\begin{figure}
\plotone{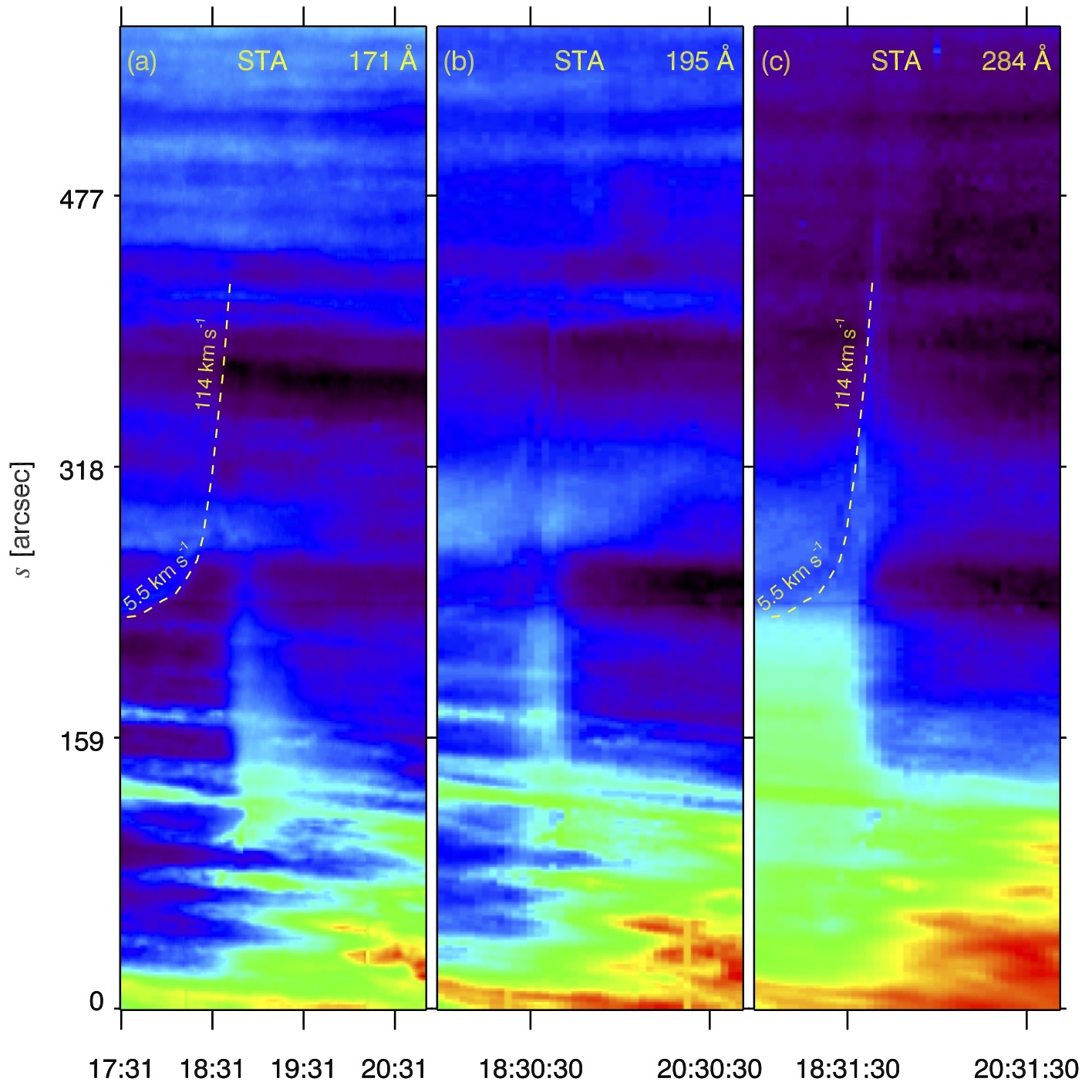}
\caption{Time-slice diagrams of S2 in 171, 195, and 284 {\AA}. $s=0$ and $s=576\farcs4$ in the $y$-axis stand for the southwest and northeast endpoints of S2.
The velocity ($\sim$114 km s$^{-1}$) of the EUV wave and the velocity ($\sim$5.5 km s$^{-1}$) of the slow expansion of CME are labeled.
\label{fig8}}
\end{figure}

In Figure~\ref{fig7}(b), we select a second slice (S2) with a length of 418 Mm, which starts from AR 11169 and extends in the same direction as the EUV wave. The time-slice diagrams of S2 in 171, 195, and 284 {\AA}
are plotted in Figure~\ref{fig8}. Like in Figure~\ref{fig6}, the bright inclined feature indicates the propagation of EUV wave in the northeast direction during 18:30$-$18:50 UT, which is overlaid by 
the yellow dashed lines.
The slopes of the dashed lines are equal to the velocity ($\sim$114 km s$^{-1}$) of EUV wave, which is close to the values in the FOV of AIA. The CME underwent a slow lateral expansion at a speed of $\sim$5.5 km s$^{-1}$ 
before $\sim$18:20 UT.

\subsection{Vertical Oscillation of the Cavity} \label{s-osci}
\begin{figure}
\plotone{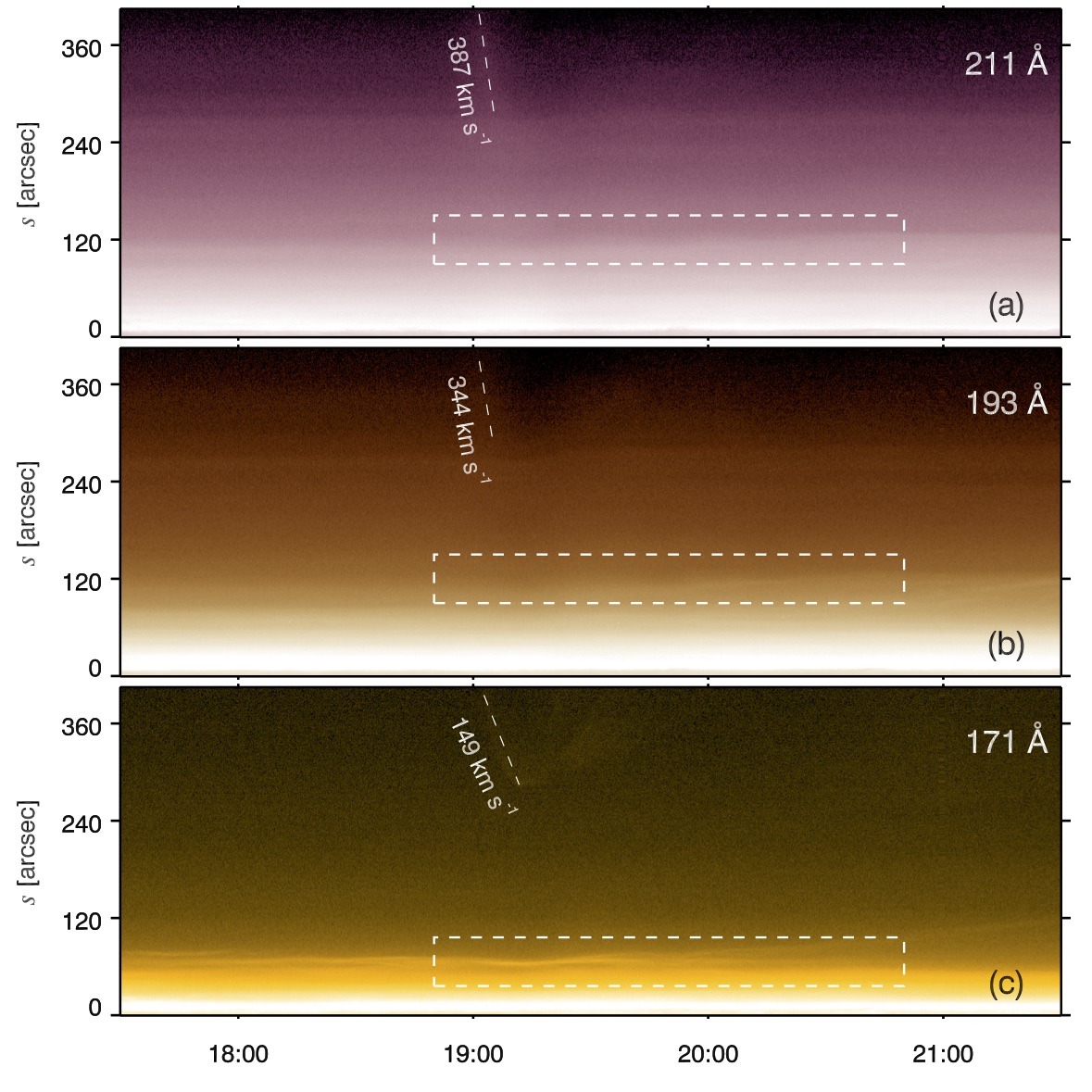}
\caption{Time-slice diagrams of S3 in 211, 193, and 171 {\AA}. $s=0$ and $s=404\farcs5$ in the $y$-axis stand for the southeast and northwest endpoints of S3.
The velocities of the downward motion of the overlying coronal loops are labeled. Close-ups of the areas within the white dashed boxes are displayed in Figure~\ref{fig10}.
\label{fig9}}
\end{figure}

From the online animation (\textit{20110316.mov}), we found clear evidence for vertical oscillation of the cavity after the arrival of the EUV wave. In Figure~\ref{fig1}(e), we select a third slice (S3).
With a length of 404$\farcs$5, S3 starts from the solar surface and passes through the cavity, covering the large-scale envelope coronal loops. 
The time-slice diagrams of S3 in 211, 193, and 171 {\AA} are plotted in Figure~\ref{fig9}. It is obvious that after the arrival of the EUV wave at $\sim$19:00 UT, the overlying coronal loops are 
pressed down before bouncing back gradually to their initial states. The velocities of the downward motion are $\sim$387, $\sim$344, and $\sim$149 km s$^{-1}$ in 211, 193, and 171 {\AA}, respectively.

\begin{figure}
\plotone{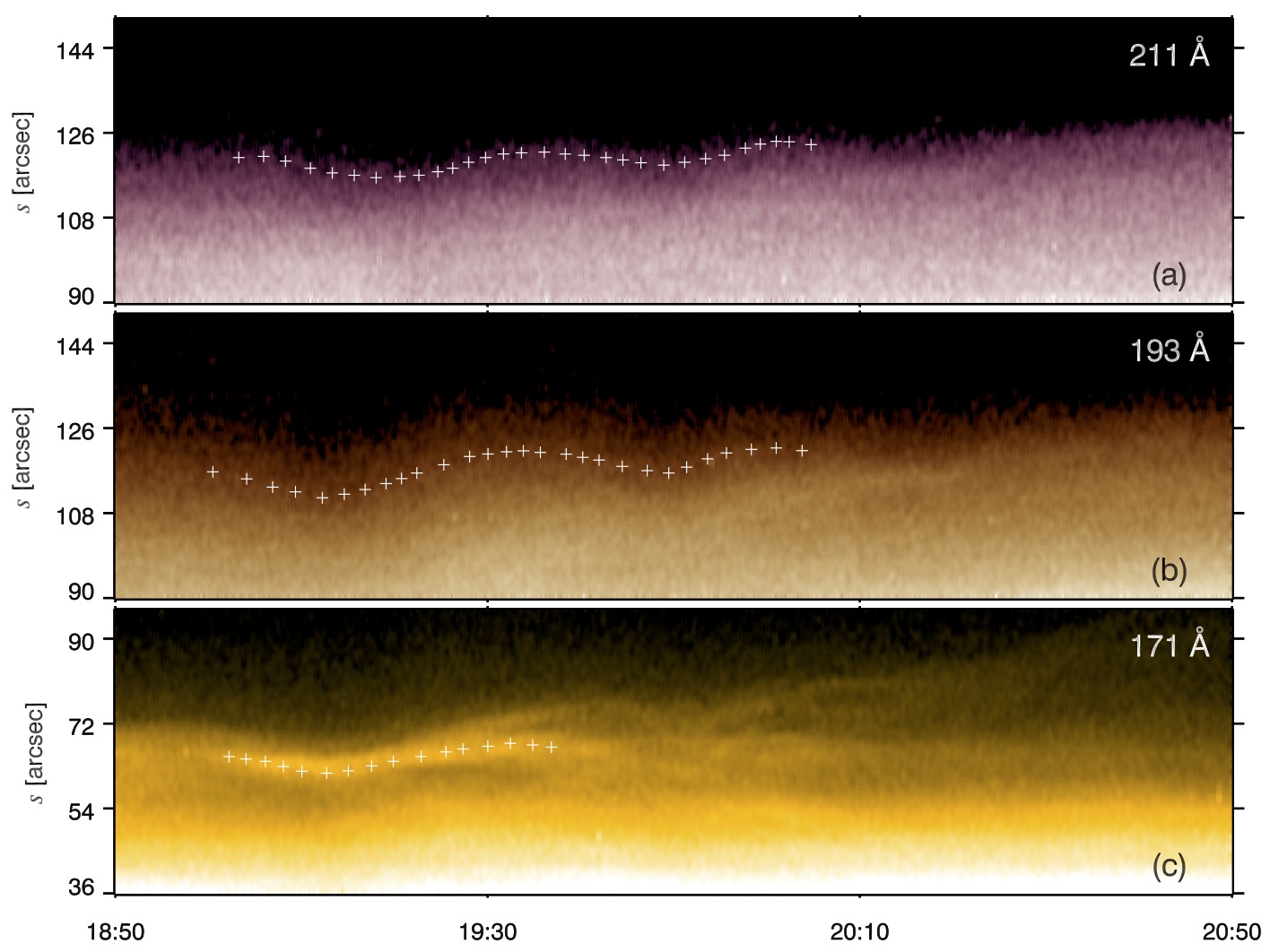}
\caption{Close-ups of the regions within the white dashed boxes of Figure~\ref{fig9}, showing vertical oscillation of the coronal cavity.
The white plus symbols are manually marked positions of the cavity.
\label{fig10}}
\end{figure}

Oscillation of the cavity with small amplitudes could not be distinctly displayed. However, when we magnify the regions within the white boxes of Figure~\ref{fig9}, things are different.
Close-ups of the regions with a better contrast are shown in Figure~\ref{fig10}. Now, the damping oscillations lasting for about 2 cycles are clear. We mark the positions of the cavity manually 
with white plus symbols. In order to obtain parameters of the oscillations, we performed curve fittings using the standard \textit{SSW} program \textit{mpfit.pro} and the same function as that 
described in previous literatures \citep{zqm12,zqm17a,zqm17b}:
\begin{equation} \label{eqn-1}
y=y_0+bt+A_0\sin(\frac{2\pi}{P}t+\phi_0)e^{-t/\tau},
\end{equation}
where $y_0$, $A_0$, and $\phi_0$ represent the initial position, amplitude, and phase. $b$, $P$, and $\tau$ stand for the linear velocity of cavity, period, and damping time of the oscillations.

\begin{figure}
\plotone{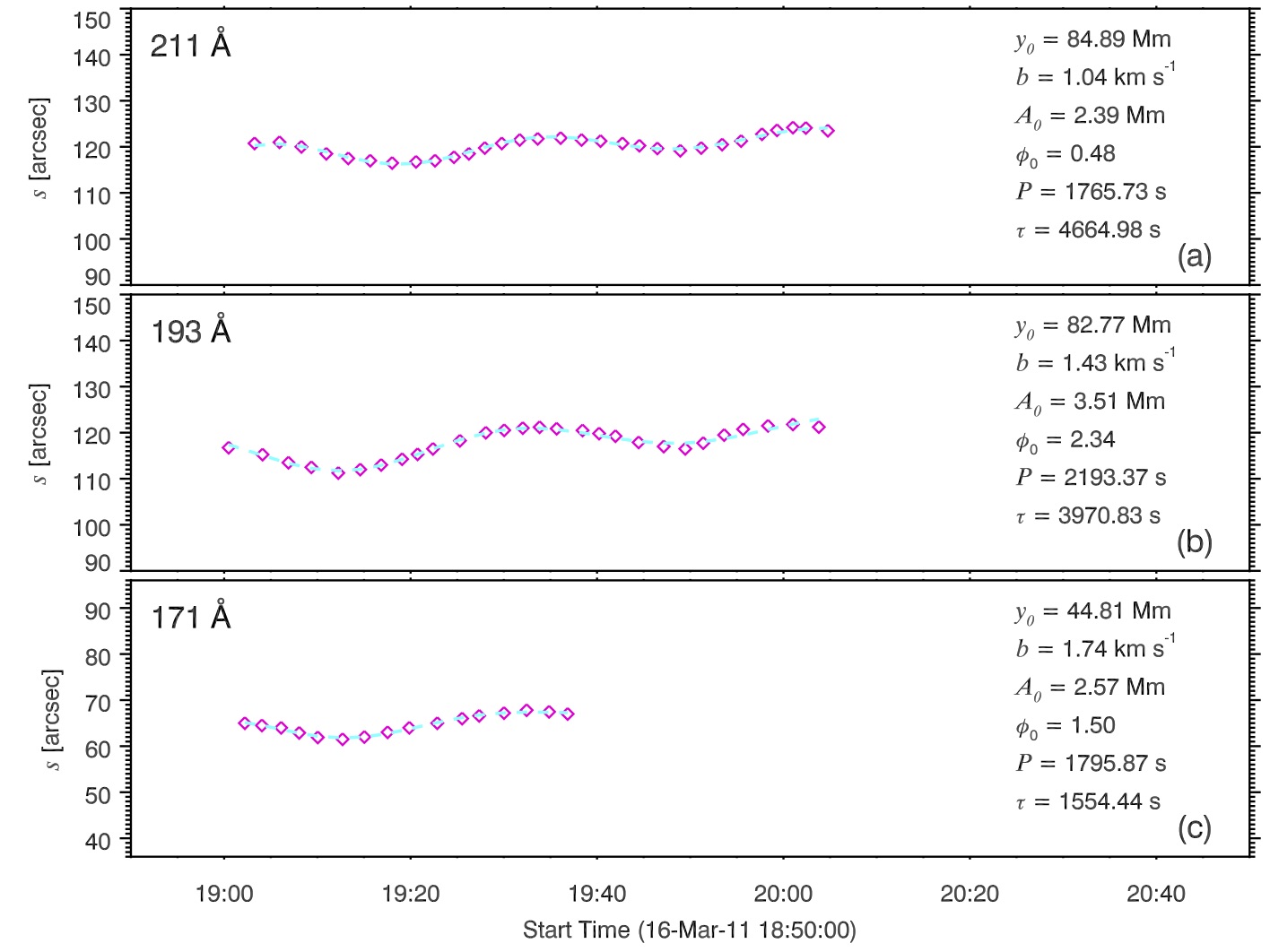}
\caption{Manually marked positions of the cavity in Figure~\ref{fig10} (magenta diamonds)
and the results of curve fittings (dashed cyan lines) in 211, 193, and 171 {\AA}.
Parameters of the oscillations are labeled.
\label{fig11}}
\end{figure}

The results of curve fittings are shown in Figure~\ref{fig11}, indicating that the function in Equation~\ref{eqn-1} can excellently describe the vertical oscillation of the cavity. 
The parameters in different wavelengths are labeled in each panel and listed in Table~\ref{tab:fitting} for a easier comparison. The amplitudes, periods, and damping times are 2.4$-$3.5 Mm,
29$-$37 minutes, and 26$-$78 minutes, respectively. The amplitudes are consistent with the values for the vertical oscillation of MFRs \citep{kim14,zhou16}. 
The periods are close to the values for the vertical oscillation of prominences \citep{hyd66,gil08,bocc11}.
Since the prominence-cavity system consists of multithermal plasma, the initial positions of oscillation in different wavelengths, which are labeled with white circles in Figure~\ref{fig1}(c-e), 
agree with their formation temperatures. In other words, hot plasmas oscillate at higher altitudes and cool plasmas oscillate at lower altitudes, implying that the prominence-carrying cavity 
oscillates as a whole body after the arrival of the EUV wave. 
The positive values of $b$ in the third column of Table~\ref{tab:fitting} suggest that the cavity ascended slowly at a speed of 1$-$2 km s$^{-1}$ during the oscillation.

\begin{figure}
\plotone{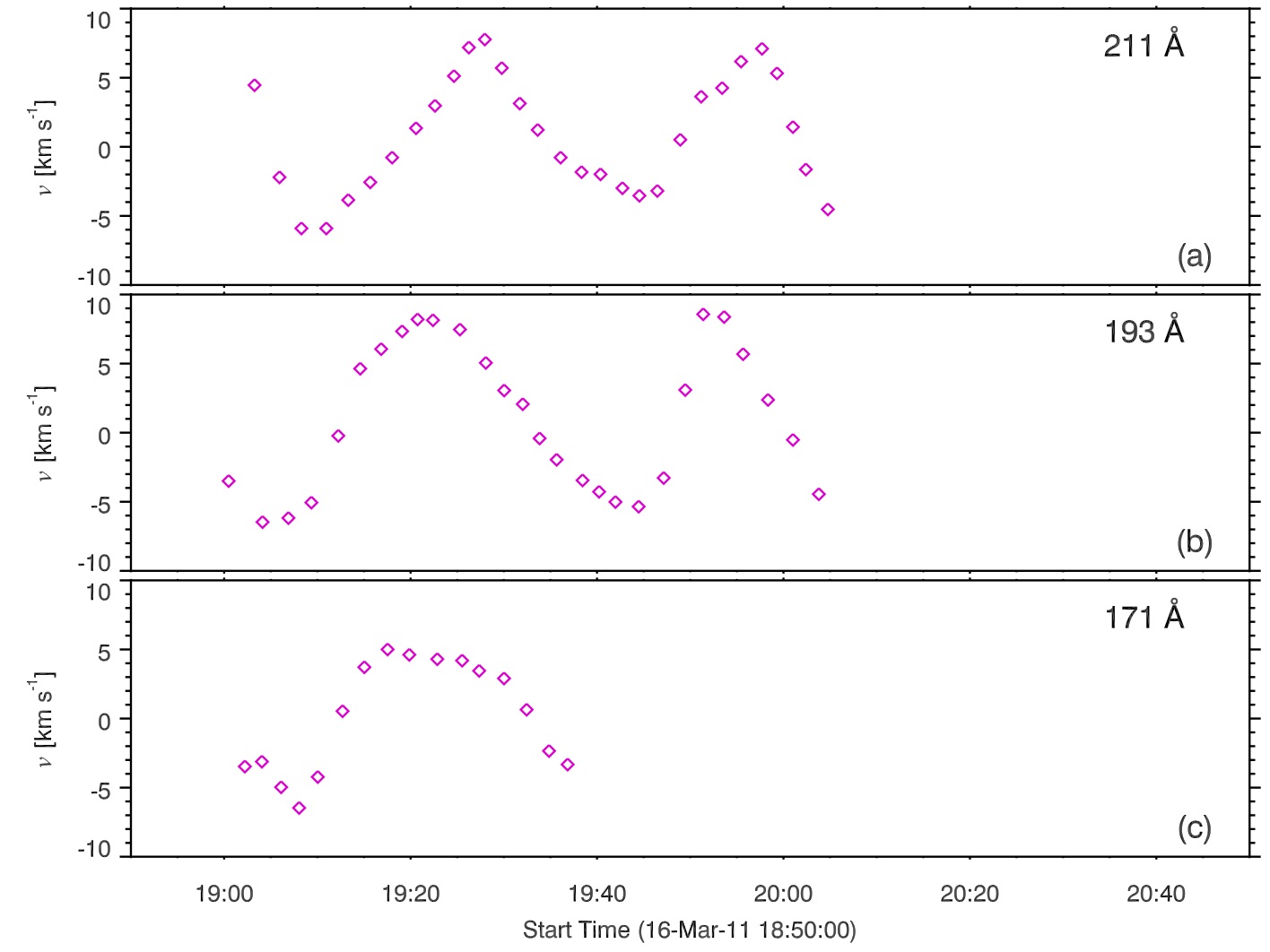}
\caption{Velocities of the vertical oscillations in 211, 193, and 171 {\AA}.
\label{fig12}}
\end{figure}

Using the simple formula $v=ds/dt$, the velocities of the vertical oscillations in 211, 193, and 171 {\AA} are calculated and displayed in Figure~\ref{fig12}. The velocity amplitudes of 
the oscillations are less than 10 km s$^{-1}$, which are close to the values previously reported by \citet{shen14a} and \citet{zhou16}. 

\begin{deluxetable}{ccccccc}
\tablecaption{Fitted parameters of the vertical oscillation \label{tab:fitting}}
\tablecolumns{7}
\tablenum{2}
\tablewidth{0pt}
\tablehead{
\colhead{$\lambda$} &
\colhead{$y_0$} & 
\colhead{$b$} &
\colhead{$A_0$} &
\colhead{$\phi_0$} &
\colhead{$P$} &
\colhead{$\tau$}\\
\colhead{({\AA})} & 
\colhead{(Mm)} &
\colhead{(km s$^{-1}$)} &
\colhead{(Mm)} &
\colhead{(rad)} &
\colhead{(min)} &
\colhead{(min)}
}
\startdata
  211 & 84.89 & 1.04 & 2.39 & 0.48 & 29.4 & 77.8 \\
 193 & 82.77 & 1.43 & 3.51 &  2.34 & 36.6 & 66.2 \\
 171 & 44.81 & 1.74 & 2.57 &  1.50 & 29.9 & 25.9 \\
\enddata
\end{deluxetable}

\section{Discussion} \label{sec:discuss}

\subsection{How is the Oscillation Triggered?}
\begin{figure}
\plotone{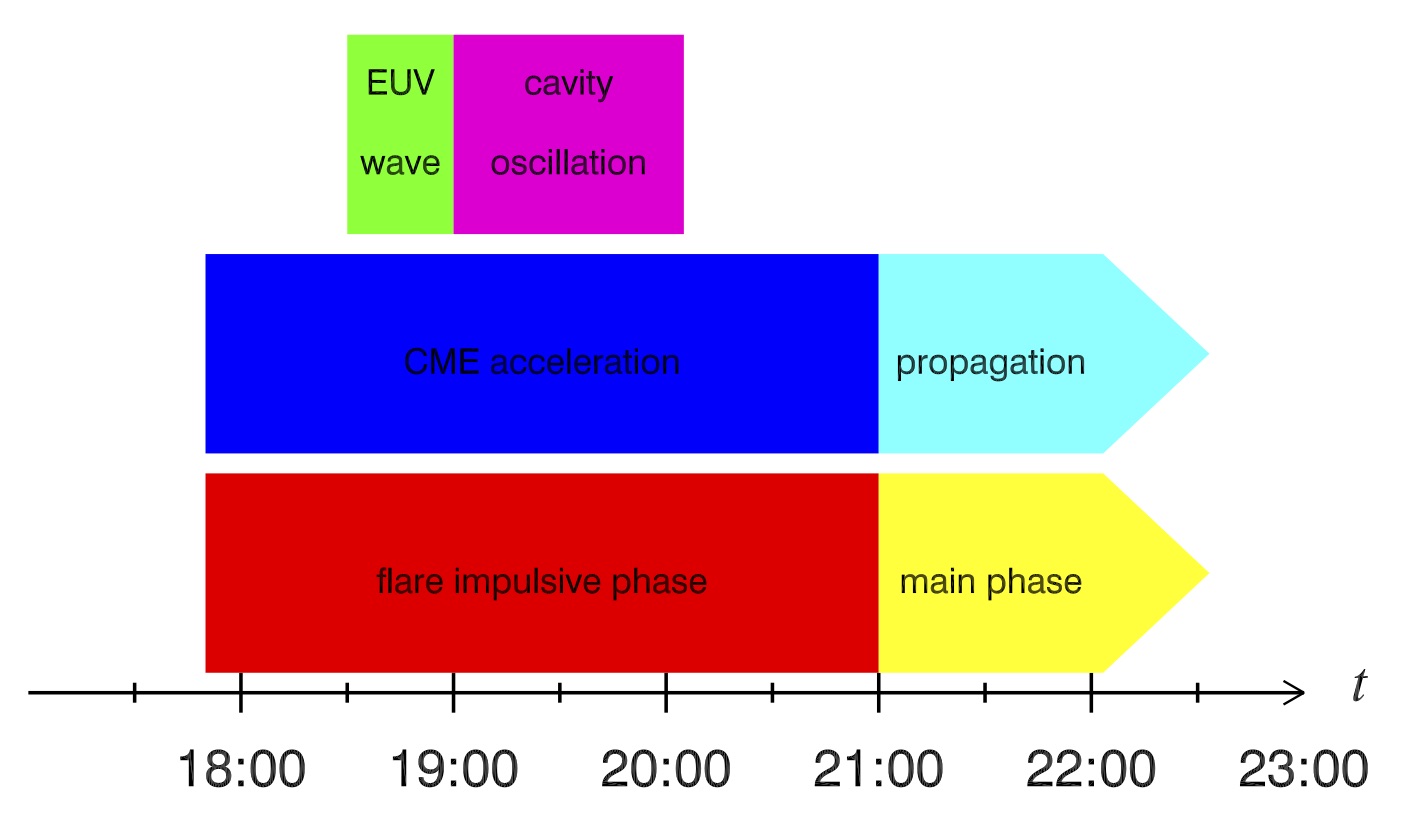}
\caption{Timeline of the whole events on 2011 March 16, including the flare, CME, EUV wave, and vertical oscillation of the cavity. 
\label{fig13}}
\end{figure}

Despite of substantial observations and investigations since the discovery of prominence oscillations \citep[e.g.,][]{hyd66,ram66}, the triggering mechanisms of prominence oscillations are far from clear.
The longitudinal oscillations can be triggered by microflares or subflares \citep{jing03,zqm12}, shock waves \citep{shen14b}, magnetic reconnections in the filament channels \citep{zqm17a}, 
and coronal jets at the legs of filaments \citep{luna14} or from a remote AR \citep{zqm17b}. The transverse oscillations, including horizontal and vertical oscillations, are usually triggered by EUV or Moreton 
waves \citep[e.g.,][]{eto02,liu12,shen14a,shen17}. \citet{liu12} reported the transverse oscillation of a cavity triggered by the arrival of a coronal EUV wave from the remote AR 11105 on 2010 September 8-9. 
The oscillation is explained by the fast kink-mode wave. In our study, the oscillation of the prominence-carrying cavity is also triggered by a coronal EUV wave from AR 11169. Timeline of the whole 
events is drawn in Figure~\ref{fig13}. On the one hand, the direction of oscillation is vertical (see Figure~\ref{fig10}), which is different from the situation in \citet{liu12}. This is most probably due to the 
different location of interplay. As shown in the schematic cartoon of \citet{liu12}, the EUV wave front collides with the cavity laterally. In our case, the EUV wave collides with the cavity from the top, which is supported 
by the fast downward motion and bouncing back of the overlying magnetic field lines above the cavity (see Figure~\ref{fig9}). Horizontal oscillation is not found from the time-slice diagrams of S1 (see Figure~\ref{fig6}).
On the other hand, the speed ($\sim$120 km s$^{-1}$) of the EUV wave from AR 11169 accounts for only 10\% of that in \citet{liu12}. This is probably due to the different component of EUV wave. 
On 2010 September 8, the fast component of the global EUV wave interacting with the cavity is interpreted by a fast MHD wave. Considering the absence of fast component EUV wave in the time-slice 
diagrams of S1 and S2 (see Figure~\ref{fig6} and Figure~\ref{fig8}), the EUV wave in our case is most probably the slow component of CME-caused restructuring \citep{chen02,chen05}.

\begin{figure}
\plotone{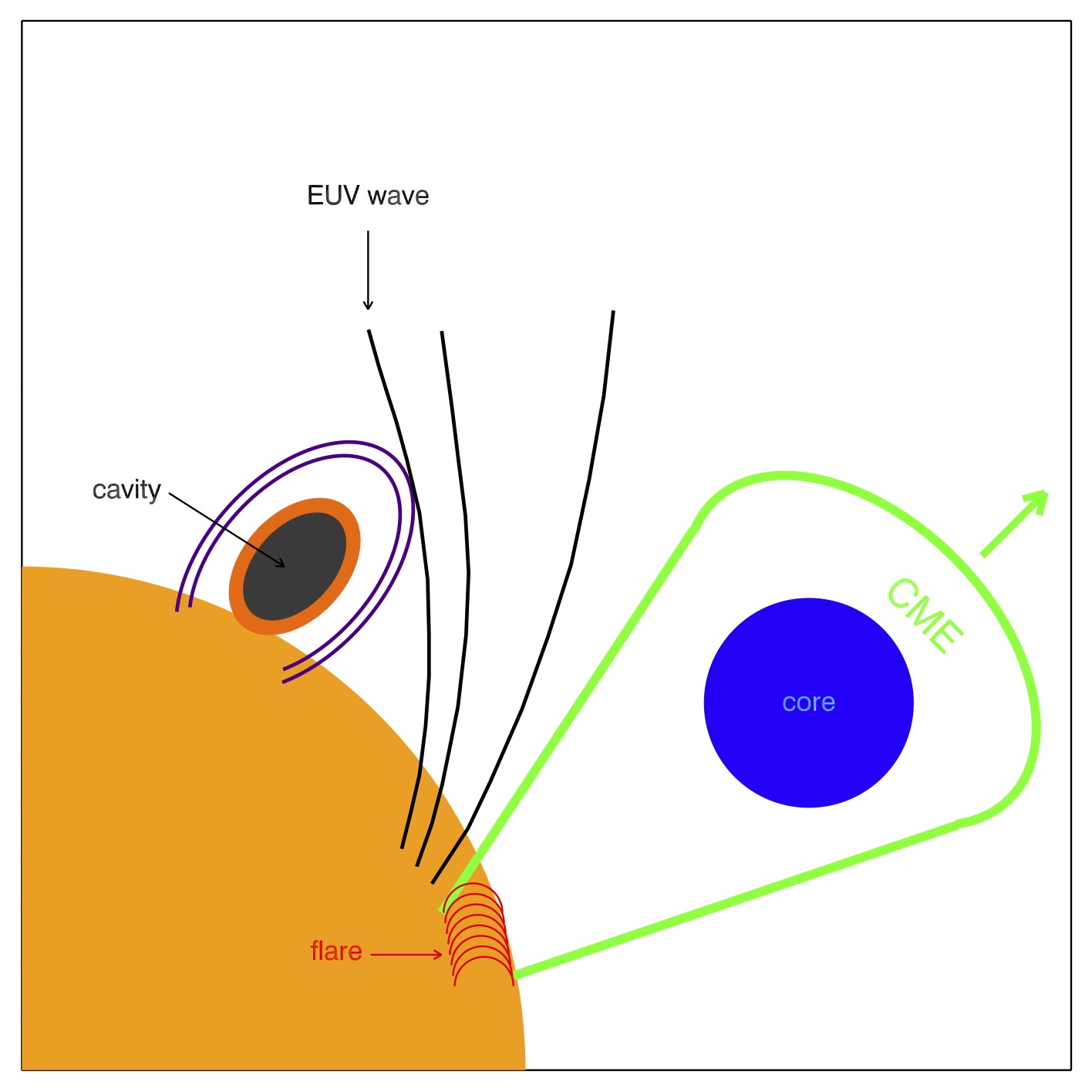}
\caption{Schematic cartoon to illustrate the interaction between a cavity and an EUV wave from a remote AR where a two-ribbon flare and a CME take place 
(see text for details). 
\label{fig14}}
\end{figure}

In Figure~\ref{fig14}, we draw a schematic cartoon to illustrate the interaction between the cavity and the EUV wave. In AR 11169 close to the solar limb, a C3.8 two-ribbon flare (red arcades) and a CME (green line 
and blue circle) take place. Owing to the stretch and lateral expansion of the magnetic field lines constraining the CME, a coronal EUV wave (black lines) is generated and propagates in the northeast direction. 
As soon as the EUV wave reaches the cavity from top, it presses the overlying magnetic field lines (purple lines) above the cavity, leading to fast downward motion of the magnetic loops. At the same time, 
the cavity (dark ellipse) and the prominence at the bottom start to oscillate in the vertical direction. Due to the quick attenuation, the oscillation lasts for $\sim$2 cycles and disappears.

\subsection{Magnetic Field Strength of the Cavity}
The restoring force of vertical prominence oscillation is another key issue that should be addressed. Recently, \citet{zhou18} performed three-dimensional (3D) ideal MHD simulations of prominence oscillations 
along a MFR in three directions. The vertical oscillation has a period of $\sim$14 minutes, which can be well explained using a slab model \citep{diaz01}. The magnetic tension force serves as the dominant 
restoring force. Considering that the periods of oscillation at different heights are approximately equal, we conclude that the cavity experienced a global vertical oscillation of fast kink mode \citep{zhou16}. 
The magnetic field strength of the cavity can be roughly estimated as follows:
\begin{equation} \label{eqn-2}
B_c=2L_c\sqrt{4\pi\rho_c}/P,
\end{equation}
where $L_c$ and $\rho_c$ represent the total length and density of the cavity, and $P$ denotes the period of oscillation. Since the length of cavity could not be measured directly, we take $L_c$ in the range of 
0.06$-$2.9 $R_{\odot}$ \citep{kar15a}. We adopt the typical density of $n_{e}=1.0\times10^8$ cm$^{-3}$ ($\rho_c=1.67\times10^{-16}$ g cm$^{-3}$) for the coronal cavity at a height of 1.13 $R_{\odot}$ \citep{ful09,sch11}.
Using the fitted period of $\sim$30 minutes (see Table~\ref{tab:fitting}), $B_c$ is calculated to be 0.2$-$10 G, which is consistent with the value of 6 G in \citet{liu12}. 

Another way of estimating the magnetic field strength of prominence at the bottom of cavity is described by \citet{hyd66}:
\begin{equation} \label{eqn-3}
B_{r}^2=\pi\rho_{p} r_{0}^2(4\pi^2 P^{-2}+\tau^{-2}),
\end{equation}
where $B_r$, $\rho_{p}$, and $r_0$ are the radial component of magnetic field, density, and scale height of the prominence, $P$ and $\tau$ are the period and damping time of the oscillation. 
Equation~\ref{eqn-3} can be simplified:
\begin{equation} \label{eqn-4}
B_{r}^2=4.8\times10^{-12}r_{0}^2(P^{-2}+0.025\tau^{-2}),
\end{equation}
if we assume $\rho_p=4\times10^{-14}$ g cm$^{-3}$, which is two orders of magnitude higher than $\rho_c$, and $r_0=3\times10^9$ cm \citep{shen17}. Using the fitted period of $\sim$30 minutes 
and damping time of 25.9 minutes (see Table~\ref{tab:fitting}), $B_r$ is calculated to be 3.8 G, which is close to the values reported in previous literatures \citep{hyd66,shen14a,shen17}.
In brief, the magnetic field strength of the prominence-carrying cavity in our case is a few Gauss and less than 10 G.

\section{Summary} \label{sec:summary}
In this paper, we report our multiwavelength observations of the vertical oscillation of a coronal cavity on 2011 March 16. The main results are summarized as follows:      
\begin{enumerate}
\item{In EUV wavelengths, the width and height of the elliptical cavity are $\sim$150$\arcsec$ and $\sim$240$\arcsec$. The centroid of the cavity has a height of $\sim$128$\arcsec$ (0.13 $R_{\odot}$) 
above the solar surface. At the bottom of the cavity, there is a horn-like quiescent prominence.}
\item{At $\sim$17:50 UT, a C3.8 two-ribbon flare and a partial halo CME took place in AR 11169 close to the western limb. The EUV and SXR intensities of the flare rose gradually to 
the peak values at $\sim$21:00 UT. The linear velocity, central position angle, and angular width of the CME are 682 km s$^{-1}$, 268$^{\circ}$, and 184$^{\circ}$ in the LASCO/C2 FOV.}
\item{During the impulsive phase of the flare and acceleration phase of the CME, a coronal EUV wave was generated and propagated in the northeast direction at a speed of $\sim$120 km s$^{-1}$. 
The EUV wave is interpreted by the field line stretching and restructuring caused by the eruption of CME.}
\item{Once the EUV wave arrived at the cavity, it interacted with the large-scale overlying magnetic field lines from the top, resulting in quick downward motion and bouncing back of the overlying loops.
Meanwhile, the cavity started to oscillate coherently in the vertical direction. The amplitude, period, and damping time are 2.4$-$3.5 Mm, 29$-$37 minutes, and 26$-$78 minutes, respectively.
The oscillation lasted for $\sim$2 cycles before disappearing.}
\item{The vertical oscillation of the cavity is explained by a global standing MHD wave of fast kink mode. Using two independent methods of prominence seismology, we carry out a rough estimation of 
the magnetic field strength of the cavity, which is a few Gauss and less than 10 G. Additional case studies using the multiwavelength and high-resolution observations are required to investigate the interaction 
between a global EUV wave and a prominence-cavity system. MHD numerical simulations are worthwhile to gain an insight into the restoring forces and damping mechanisms of transverse prominence oscillations.}
\end{enumerate}

\acknowledgments
We would like to thank Y. N. Su, D. Li, T. Li, R. S. Zheng, and Y. D. Shen for fruitful and valuable discussions. 
\textit{SDO} is a mission of NASA\rq{}s Living With a Star Program. AIA and HMI data are courtesy of the NASA/\textit{SDO} science teams. 
This work utilizes GONG data from NSO, which is operated by AURA under a cooperative agreement with NSF and with additional financial support from NOAA, NASA, and USAF.
QMZ is supported by the Youth Innovation Promotion Association CAS, NSFC (No. 11333009, 11790302, 11773079), the Fund of Jiangsu Province (BK20161618 and BK20161095), 
``Strategic Pilot Projects in Space Science'' of CAS (XDA15052200), and CAS Key Laboratory of Solar Activity, National Astronomical Observatories (KLSA201716).

\end{document}